\documentclass[runningheads]{cl2emult}

\usepackage{makeidx}  
\usepackage{graphicx} 
\usepackage{subeqnar} 
\usepackage{multicol} 
\usepackage{cropmark} 
\usepackage{lnp}      
\makeindex            

\begin{document}
\title*{Shell Models for Magnetohydrodynamic \protect\newline 
Turbulence}
\toctitle{Shell Models for Magnetohydrodynamic
\protect\newline Turbulence}
%
%
\titlerunning{Shell Models for MHD}
%
\author{Paolo Giuliani}
%
\authorrunning{Paolo Giuliani}
%
%
\institute{Department of Physics, University of Calabria,
87036 Rende (CS), Italy}
\maketitle              

\begin{abstract}
We review the main properties of shell models for magnetohydrodynamic (MHD)
turbulence. After a brief account on shell models with nearest neighbour
interactions, the paper focuses on the most recent results
concerning dynamical properties and intermittency
of a model which is a generalization to MHD of the
Gledzer-Yamada-Okhitani (GOY) model for hydrodynamic. 
Applications to astrophysical problems are also discussed.
\end{abstract}

\section{Introduction}
Shell models are dynamical systems (ordinary differential equations)
representing a simplified version of the spectral Navier--Stokes or MHD
equations for turbulence. They were originally introduced and developed bydfg
Obukhov~\cite{Ob}, Desnyansky and Novikov \cite{DN} and Gledzer \cite{Gl}
in hydrodynamic turbulence and constitute nowdays a consistent and
relevant alternative approach
to the analytical and numerical study of fully developed turbulence
(see~\cite{BJPV} for a complete review).

Shell models are built up by
dividing wave-vector space (k--space) in a discrete number of shells
whose radii grow exponentially like $k_n=k_0\lambda^n$, ($\lambda > 1$),
$n=1\mbox{, }2\mbox{,}\ldots\mbox{, }N$. Each shell is assigned a scalar
dynamic variable, $u_n(t)$, (real or complex) which takes into account
the averaged effects of velocity modes between $k_n$ and $k_{n+1}$. The
equation for $u_n(t)$ is then written in the formex
\begin{equation}
\frac{du_n}{dt} = k_nC_n+D_n+F_n
\end{equation}
where $k_nC_n$, $D_n$ and $F_n$ are respectively quadratic nonlinear coupling
terms (involving nearest or next-nearest shell interactions), dissipation
terms and forcing
terms, the last 
generally restricted to the first shells. Nonlinear terms are chosen
to satisfy scale-invariance and conservation of ideal invariants. The main
advantage shell models offer is that they can be investigated by means of
rather easy numerical simulations at very high Reynolds ($Re$) numbers.
The degrees of freedom of a shell model are $N\sim\ln Re$, to be compared
with $N\sim Re^{9/4}$ for a three dimensional hydrodynamic turbulence following
the Kolmogorov scaling.  

The paper is organized as follows. In section 2 shell models with
nearest neighbour interactions are briefly reviewed. In section 3
equations for MHD models with nearest and next-nearest neighbour
interactions are presented and conservations laws for the ideal case
are discussed. Section 4 is devoted to dynamo action in shell models
and section 5 to spectral properties in forced stationary state and
intermittency. In section 6 conclusions are drawn and a brief mention to
astrophysical applications is made.

\section{Models with nearest neighbour interactions}
The simplest hydrodynamic shell model is the Obukhov--Novikov model,
which is a linear superposition of the Obukhov equation \cite{Ob} and
the Novikov equation \cite{DN}.
The model involves real variables $u_n(t)$ and conserves
the energy $1/2\sum_{n=1}^{N}u_n^2$ in absence
of forcing and dissipation. It does not conserve phase space volume
nor other quadratic invariants exist. The extension of the Obukhov-Novikov 
model to MHD is due to Gloaguen {\it et al.}~\cite{GLPG}. We write
down the equations for clarity
($u_n$ and $b_n$ represent 
respectively the velocity and the magnetic field in dimensionless units)
\begin{eqnarray}
\label{previousmodel1}
{du_n \over dt} = -\nu k_{n}^{2}u_{n} 
		+\alpha\left(k_{n}u_{n-1}^2-k_{n+1}u_{n}u_{n+1}-k_{n}b_{n-1}^2
                + k_{n+1}b_{n}b_{n+1}\right) \\
                + \beta\left(k_n u_{n-1}u_{n}-k_{n+1}u_{n+1}^2-k_n b_n b_{n-1}
                + k_{n+1}b_{n+1}^2\right) \nonumber
\end{eqnarray}
\vglue -0.85 truecm
\begin{eqnarray}
\label{previousmodel2}
{db_n \over dt} = -\eta k_{n}^{2}b_{n}
		+ \alpha k_{n+1}\left(u_{n+1}b_n -u_n b_{n+1})
		+ \beta k_n ( u_n b_{n-1} - u_{n-1}b_{n}\right)
\end{eqnarray}
Here $\nu$ is the kinematic viscosity, $\eta$ is the magnetic diffusivity,
$\alpha$ and $\beta$ are two arbitrary coupling coefficients.
The ideal invariants of the system are the total energy, 
$1/2\sum_{n=1}^{N}\left(\,u_n^2+b_n^2\,\right)$ and the cross-correlation,
$\sum_{n=1}^{N}u_nb_n$ which are two ideal invariants of the MHD equations
\cite{BiB}. When written in terms of the Els\"asser variables
$Z_n^+=u_n+b_n$, $Z_n^-=u_n-b_n$, the equations assume a simmetric form
and the conservation of the two previous invariants is equivalently expressed
as the conservation of the pseudo-energies
$E^{\pm}=(1/4)\sum_{n=1}^{N}\,Z_n^{\pm}\,^2$. It is remarkable to note
that, unlike the hydrodynamic model,
the MHD version satisfies a Liouville theorem
$\sum_{n=1}^{N}\partial({dZ_n^{\pm}/dt})/\partial Z_n^{\pm}=0$,
impling phase-space volume conservation.
The MHD equations conserve a third ideal invariant which is the magnetic
helicity
in three dimensions (3D) and the mean square potential in two dimensions (2D),
but no further ideal quadratic invariant can be imposed to this shell model.

A detailed bifurcation
analysis for a three-mode system was performed in \cite{GLPG}
for different
values of $\alpha$ and $\beta$. The low Reynolds (kinetic and magnetic)
numbers, used as control parameters,
allowed to identify
a great variety of regions in the
parameter space. Turbulence was investigated with a nine-mode system which
produces an inertial range with spectra following approximately
the Kolmogorov scaling $E(k)\sim k^{-5/3}$.
Temporal intermittency was also observed
and then reconsidered in more details by
Carbone \cite{Vincenzo1} who calculated the scaling exponents
of the structure functions for the Els\"asser variables and for the
pseudo-energy transfer rates, showing consistency
with the usual multifractal theory. Other interesting MHD phenomena
were also observed in \cite{GLPG} such as dynamo effect and the growth
of correlation between velocity and magnetic field in an unforced
simulation. These phenomena will be treated in more details in the
next paragraphs. 

The complex version of (\ref{previousmodel1}) and
(\ref{previousmodel2}) was
thoroughly investigated by Biskamp \cite{Bip}. The complex model allows
to include the Alfv\`en effect \cite{I}, \cite{Kr}, \cite{DMV},
that is the interaction of a constant
large scale magnetic field with small scale turbulent eddies. The main
consequence of this effect should be a reduction of the spectral energy
transfer rate and a consequent change of the spectra from the 
Kolmogorov scaling, $E(k)\sim\,k^{-5/3}$, to the Iroshnikov-Kraichnan one,
$E(k)\sim k^{-3/2}$. In this paper the Alfv\`en effect
will not be furtherly treated. The reader is referred to
\cite{Bip} for a complete discussion concerning the inclusion of
Alfv\`enic terms in shell models.

\section{Models with nearest and next-nearest interactions}
Shell models with nearest and next-nearest neighbour interactions
were introduced
by Gledzer \cite{Gl}. In particular the so called GOY
(Gledzer-Yamada-Ohkitani) model has
been extensively both numerically and analitically investigated \cite {YO},
\cite{JPV}, \cite{BLLP}.
The GOY model allows to conserve another quadratic invariant besides energy
which was identified with the kinetic helicity \cite{KLBW}.
A generalization of the GOY model to MHD can be found in
Biskamp \cite{Bip}. All the parameters of the model are now
fixed by imposing the conservation of another quadratic invariant
that can be chosen to distinguish between a 3D and a 2D model. A more 
refined version
was then considered by Frick and Sokoloff \cite {FS} to take into account
the fact that the magnetic helicity is a quantity not positive definite.
The situation can be summarized as follows \cite{PeV}.

Let us consider the following set of equations ($u_n$ and $b_n$
are now complex
variables representing the velocity and the magnetic field 
in dimensionless units)
\begin{eqnarray}
{du_n \over dt} = -\nu k_{n}^{2}u_{n}-\nu^{\prime} k_{n}^{-2}u_{n}+ik_{n}\Bigl\{
(u_{n+1}u_{n+2}-b_{n+1}b_{n+2}) \nonumber \\
- \frac{\delta}{\lambda}(u_{n-1}u_{n+1}-b_{n-1}b_{n+1})
- \frac{1-\delta}{\lambda^2}(u_{n-2}u_{n-1}-b_{n-2}b_{n-1})
\Bigr\}^{\ast}+f_{n}
\label{nonlinearev}
\end{eqnarray}
\begin{eqnarray}
{db_n \over dt} = -\eta k_{n}^{2}b_{n}+ik_n\Bigl\{
(1-\delta-\delta_m)(u_{n+1}b_{n+2}-b_{n+1}u_{n+2})\nonumber \\
+\frac{\delta_m}{\lambda}(u_{n-1}b_{n+1}-b_{n-1}u_{n+1})+
\frac{1-\delta_m}{\lambda ^{2}}(u_{n-2}b_{n-1}-b_{n-2}u_{n-1})
\Bigr\}^{\ast}
+g_n
\label{nonlineareb}
\end{eqnarray}
or, in terms of the complex 
Els\"asser variables $Z_n^{\pm}(t) = v_n(t) \pm b_n(t)$, particularly
useful in some solar--wind applications,
\begin{equation}
\label{nonlineare}
{dZ_n^{\pm} \over dt}=-\nu^{+}k_{n}^{2}Z_n^{\pm}
-\nu^{-}k_{n}^{2}Z_n^{\mp}
-\frac{\nu^{\prime}}{2}k_n^{-2}Z_n^{+}
-\frac{\nu^{\prime}}{2}k_n^{-2}Z_n^{-}
+ik_{n}T_{n}^{\pm *}+f_{n}^{\pm}
\end{equation}
where
\begin{eqnarray}
T_n^{\pm}             & = & \left\{  
{{\delta +\delta_{m}}\over 2}
Z_{n+1}^{\pm}Z_{n+2}^{\mp}+{{2-\delta -\delta_{m}}\over 2}
Z_{n+1}^{\mp}Z_{n+2}^{\pm} \right. \nonumber \\
                      &   &
\mbox{}+{{\delta_{m}-\delta}\over {2\lambda}}
Z_{n+1}^{\pm}Z_{n-1}^{\mp}-{{\delta +\delta_m}\over {2\lambda}}Z_{n+1}
^{\mp}Z_{n-1}^{\pm} \nonumber \\
                      &   &
\left. \mbox{}-{{\delta_{m}-\delta}\over{2\lambda^{2}}}
Z_{n-1}^{\pm}Z_{n-2}^{\mp}-
{{2-\delta -\delta_{m}}\over {2\lambda^{2}}}Z_{n-1}^{\mp}Z_{n-2}^{\pm}
\right\}
\end{eqnarray}
Here $\nu^{\pm}=(\nu\pm\eta)/2$,
being $\nu$ the kinematic viscosity and $\eta$ the resistivity,
$-\nu^{\prime} k_{n}^{-2}u_{n}$, eq. (\ref{nonlinearev}), is a drag term
specific to 2D cases (see below),
$f_n^{\pm}=(f_n\pm g_n)/2$ are external driving forces,
$\delta$ and $\delta_m$ are real
coupling coefficients to be determined.
In the inviscid unforced limit, equations (\ref{nonlineare}) conserve both
pseudoenergies
$
E^{\pm}(t) = (1/4)\sum_n \left|Z_n^{\pm}(t)\right|^2
$
for any value of $ \delta $ and $ \delta_{m} $ 
(the sum is extended to all the shells), which corresponds to the conservation 
of both the total energy 
$E = E^{+}+E^{-}=(1/2)\sum_n (\left|v_n(t)\right|^2+
\left|b_n(t)\right|^2)$
and the cross-helicity 
$h_C =  E^{+}-E^{-}=\sum_n Re(v_n b_n^*)$.
As far as the third ideal invariant
is concerned,
we can define a generalized quantity
as
\begin{equation}
H_B^{(\alpha)}(t)=\sum\limits_{n=1}^{N} (\mbox{sign}(\delta-1))^n
\frac{|b_n(t)|^{2}}{k_n^{\alpha}}
\end{equation}
whose conservation implies $\delta = {1-\lambda^{-\alpha}}$,
$\delta_m =\lambda^{-\alpha}/(1+\lambda^{-\alpha})$ for
$\delta <1$, $0 <\delta_m<1$ and
$\delta   = {1+\lambda^{-\alpha}} $,
$\delta_m = -\lambda^{-\alpha}/(1-\lambda^{-\alpha})$ for
$\delta >1$, $\delta_m<0$, $\delta_m>1$. Thus two classes of
MHD GOY models can be defined with respect to the values of
$\delta$: 3D--like models for $\delta<1$, where $H_B^{(\alpha)}$
is not positive
definite and represents a generalized magnetic helicity;
2D--like models where $\delta>1$ and $H_B^{(\alpha)}$ is positive 
definite.
This situation strongly resembles what happens in the hydrodynamic
case where 2D--like $(\delta>1)$ and 3D--like $(\delta<1)$ models are
conventionally distinguished with
respect to a second generalized conserved quantity
$H_K^{(\alpha)}(t)=
\sum_n {(\mbox{sign}(\delta-1))^n}k_n^{\alpha} \left|v_n(t)\right|^2.$
Here the 3D and 2D cases are recovered for $\alpha=1,\,2$ where the ideal
invariants are identified respectively with kinetic helicity and enstrophy. It
should be noted that, although the hydrodynamic invariants are not conserved
in the magnetic case, the equations which link $\alpha$ and $\delta$ are
exactly the same for hydrodynamic and MHD models. Thus, once fixed $\alpha$
and $\delta$, it is a simple matter to find out
which GOY model the MHD GOY one reduces to when $b_n=0$ \cite{PeV}. 
To summarize we have that (with $\lambda=2$) the   
model introduced in \cite{FS} for the 3D case will be called, 
hence on, 3D MHD GOY model or simply 3D model. It is recovered for
$\alpha=1$, $\delta=1/2$,
$\delta_m=1/3$ and reduces to the usual 3D GOY model for $b_n=0$.
The Biskamp's 3D model \cite{Bip} is
actually a 2D--like model and will be called pseudo 3D model.
It is obtained for $\alpha=1$, $\delta=3/2$,
$\delta_m=-1$ and reduces to a 2D--like GOY model that conserves a quantity
which has the same dimensions as kinetic helicity but is positive definite.
The 2D models introduced in \cite{Bip} and in \cite{FS}
coincide, they are recovered for $\alpha=2$,
$\delta=5/4$, $\delta_m=-1/3$
and reduce to the usual 2D GOY model for $b_n=0$. In the following
the properties of the 3D model will be mainly investigated.\\

\section{Dynamo action in MHD shell models}
The problem of magnetic dynamo, that is the amplification of a seed of
magnetic field and its maintenance against the losses of dissipation in
an electrically conducting flow, is of great interest by itself and
for astrophysical applications (see for example \cite{Moffat} for
an excellent introduction to the problem). Shell models offer
the opportunity to test with relative simplicity whether a small value
of the magnetic field can grow in absence of forcing
terms on the magnetic field.
Previous considerations about dynamo action
in shell models can be found in \cite{GLPG}. In that case numerical
study of bifurcations in the three-mode system revealed
instabilities of kinetic
fixed points to magnetic ones or magnetic chaos. The existence
of a sort of dynamo effect in MHD GOY models was put forward
by Frick and Sokoloff \cite{FS}.
The authors investigate
the problem of 
the magnetic field generation in a free-decaying turbulence, 
thus showing that: $1)$ in the $3D$ case magnetic energy grows and
reaches a value comparable with the kinetic one, in a way that the magnetic 
field growth
is unbounded in the kinematic case; $2)$ in the $2D$ case magnetic energy
slowly decays in the nonlinear as well as in the kinematic case. These results
have been interpreted as a 3D ``turbulent dynamo effect" and seem to 
be in agreement with well-known results by which dynamo effect is not
possible in two dimensions \cite{Z}. The problem was then reexamined
in \cite{PeV} in a forced situation looking at a comparison between
the 3D MHD GOY model and the pseudo 3D model. 

\begin{figure}
\centering
\includegraphics[width=.7\textwidth]{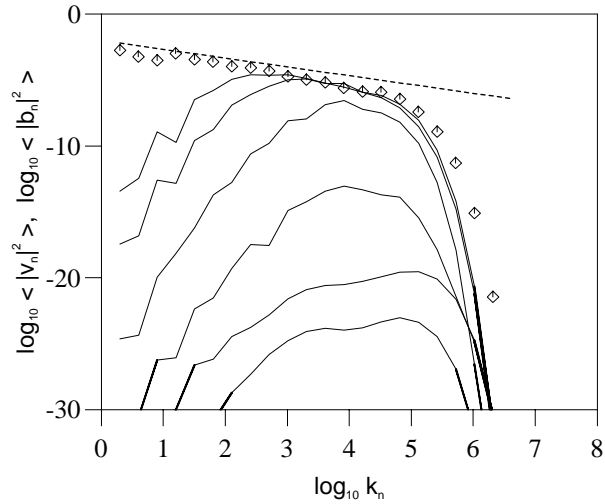}
\caption[]{3D model: $\log_{10} \langle\, |v_n|^{2} \,\rangle$ ({\it diamonds})
and $\log_{10}\langle\, |b_n|^{2}\,\rangle$ ({\it lines}) versus
$\log_{10}k_n$. 
The averages of $|b_n|^{2}$ are made over intervals of 3 large scale turnover 
times. Time proceeds upwards. The kinetic
spectrum is averaged over 30 large scale turnover times. The straight line has
slope $-2/3$. Parameters used: N=24, $\nu=\eta=10^{-8}$, $\nu^{\prime}=0$}
\label{fseps}
\end{figure}

\begin{figure}
\centering
\includegraphics[width=.7\textwidth]{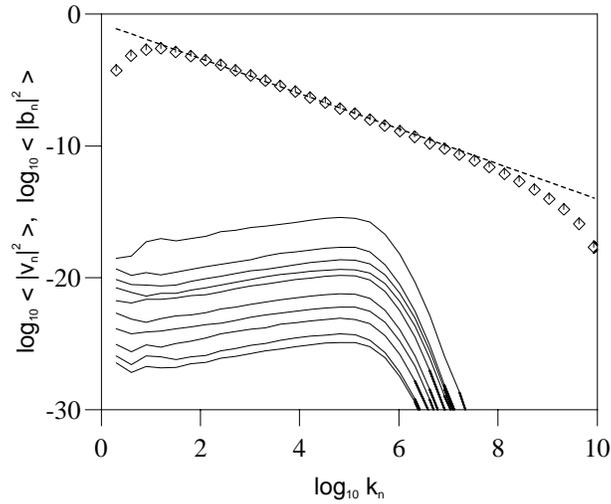}
\caption[]{Pseudo 3D model: $\log_{10} \langle\,|v_n|^{2}\,\rangle$
({\it diamonds})
and $\log_{10}\langle\,|b_n|^{2}\,\rangle$ ({\it lines}) versus 
$\log_{10}k_n$.
Averages are made over intervals of 100 large scale turnover times.
Time proceeds downwards. The kinetic
spectrum is only shown for the last interval. The straight line has
slope $-4/3$, see text for explanation. Parameters used: N=33, 
$\nu=10^{-16},\;\eta=0.5\cdot 10^{-9}$, $\nu^{\prime}=1$}
\label{biskeps}
\end{figure}

Starting from a well developed turbulent
velocity field, a seed of magnetic field is injected and the growth
of the magnetic spectra monitored. System is forced on the shell
$n=4$ ($k_0=1$), 
setting $f_4^{+}=f_4^{-}=(1+i)\,10^{-3}$, which corresponds to
only inject kinetic energy at large scales. Method of integration is a
modified fourth order Runge-Kutta scheme. 
In fig. \ref{fseps} we plot 
$\log_{10}\langle\,|b_n|^2\,\rangle$ and $\log_{10}\langle\,|v_n|^2\,\rangle$
versus $\log_{10}k_n$ for the 3D model. Angular brackets $\langle\,\,\,\rangle$
stand for
time averages. It can be seen that the magnetic energy 
grows rapidly in time and forms a spectrum where the amplitude of the various 
modes is, at small scale, of the same order as the kinetic energy spectrum.
(The subsequent evolution of magnetic
and kinetic spectra will be considered in the next section).
The spectral index is close
to $k^{-2/3}$ which is compatible with a Kolmogorov scaling of
the second order structure function.
For a comparison we integrated the pseudo 3D model
and it can be seen (fig. \ref{biskeps}) that a magnetic spectrum is
formed, but it slowly decays
in time.
Notice that, because of the smallness of $b_n$, its back-reaction on the
velocity field is negligible, thus the kinematic part of the model
evolves independently from the magnetic one.
Now the scaling $|v_n|^2\sim k_n^{-4/3}$ follows, 
as a cascade of generalized enstrophy is expected for 2D--like hydrodynamic
GOY models when $\alpha<2$ (see \cite{DM} for details).
The question now
arises whether it is correct the interpretation of the growth of the magnetic
field in the 3D model as the corresponding dynamo
effect expected in the real 3D magnetohydrodynamics.  First of all
it should be noted that in the kinematic case an analogy with the
vorticity equation predicts the following relations between velocity and
magnetic energy spectra \cite{BiB}:
$|v_n|^{2}\sim\,k^{-a},\,|b_n|^{2}\sim\,k^{2-a}$,
so that
if $a=2/3$ it follows a magnetic energy spectrum growing with $k$.
The kinematic case corresponds to the first stage of growth of our simulation
where this
behaviour is sometimes visible, at least qualitatively. Note however that
the averages are made on very small time intervals because of the rapid growth
of the magnetic energy. A similar, much more pronounced behaviour is found for
the  pseudo 3D model as
well.

Let us stress that the sign of the third ideal invariant 
seems to play a crucial role as far as the growth of small magnetic fields is 
concerned. In effect this sort of dynamo effect can also be considered under a
different point of view. Let us consider the ideal evolution
of the model 
$dZ_n^{\pm}/dt = i k_n T_n^{\pm\ast}$. We can build up the phase space $S$ of 
dimension $D = 4N$, by 
using the Els\"asser variables as axes, so that a point in $S$ represents the 
system at a given time. A careful analysis of (\ref{nonlineare}) shows 
that there exist some subspaces $I \subset S$ of dimension $D=2N$
which remain invariant under 
the time evolution \cite{CV}. More formally, let $y(0)=(v_n,b_n)$ be a set of 
initial conditions such that $y(0) \in I$, $I$ is time invariant if the 
flow $T^t$, representing the time evolution operator in $S$, leaves 
$I$ invariant, that is $T^{t}[y(0)] = y(t) \in I$.
The kinetic subspace $K \subset S$, defined by $y(0) = (v_n,0)$ is
obviously the usual fluid GOY model. 
Further subspaces are the Alfv\'enic subspaces $A^{\pm}$ defined by 
$y(0)=(v_n,\pm v_n)$, say $Z_n^+ \not = 0$ and $Z_n^-=0$ (or vice versa). Each
initial condition in these subspaces is actually a fixed point
of the system. We studied the properties of stability of $K$ and $A^{\pm}$.
Following \cite{CV}, 
let us define for each $I$ the orthogonal complement $P$, namely
$S = I \oplus P$. Let us then decompose the solution as
$y(t)=(y_{int}(t),y_{ext}(t))$ where the subscripts refer to the $I$ and
$P$ subspaces respectively. Finally we can define the energies
$E_{int}=\|y_{int}\|^2$ and $E_{ext}=\|y_{ext}\|^2$. Note that the distance
of a point $y=(y_{int},y_{ext})$ from the subspace $I$ is
$d=\min\limits_{\hat{y} \in I}\|y-\hat{y}\|=\|y_{ext}\|$.
Then $E_{ext}$ represents the square of the distance 
of the solution from the invariant subspace. At time $t=0$, $E_{ext} = 
\epsilon E_{int}$ ($\epsilon \ll 1$) represents the energy of the 
perturbation. Since the total energy is constant in the ideal case, two
extreme situations can arise: 
$1)$ The external energy remains of the same order of its initial value, that 
is the solution is trapped near $I$ which is then a stable subspace; $2)$ The 
external energy assumes values of the same order as the internal energy, 
that is the solution is repelled away from the subspace which is then unstable.%
\begin{figure}
\centering
\includegraphics[width=.7\textwidth]{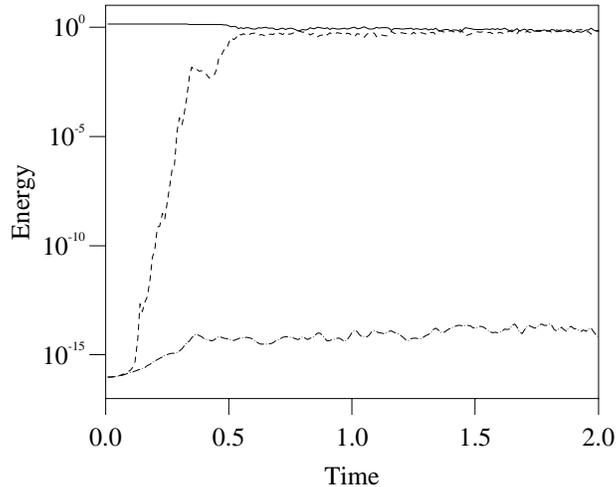}
\caption[]{Ideal case: kinetic energy ({\it continuous line}) and
magnetic energy ({\it dashed line}) versus time for the 3D model; magnetic energy
({\it dot-dashed line}) versus time for the pseudo 3D model}
\label{eidealeps}
\end{figure}
Since the external and internal energies 
for the Alfv\'enic subspaces are nothing but the pseudoenergies $E^+$ and 
$E^-$, which are ideal invariants, the Alfv\'enic subspaces are stable. As 
regards the kinetic subspace, $E_{int}$ and $E_{ext}$ represent respectively
the 
kinetic and magnetic energies. Looking at the numerical solutions of the 
ideal model (fig. \ref{eidealeps}) we can see the difference in the stability 
properties between the pseudo 3D model and the 3D one. In the first case the 
external energy remains approximately constant, while in the second case the 
system fills up immediately all the available phase space. This striking 
difference is entirely due to the nonlinear term, and in fact must be 
ascribed to the differences in sign of the third invariant.
The effect of the unstable subspace, which pushes away the solutions,
is what in ref. \cite{FS} is called ``turbulent dynamo effect".

\section{Spectral properties in stationary forced state}
\begin{figure}
\centering
\includegraphics[width=1.\textwidth]{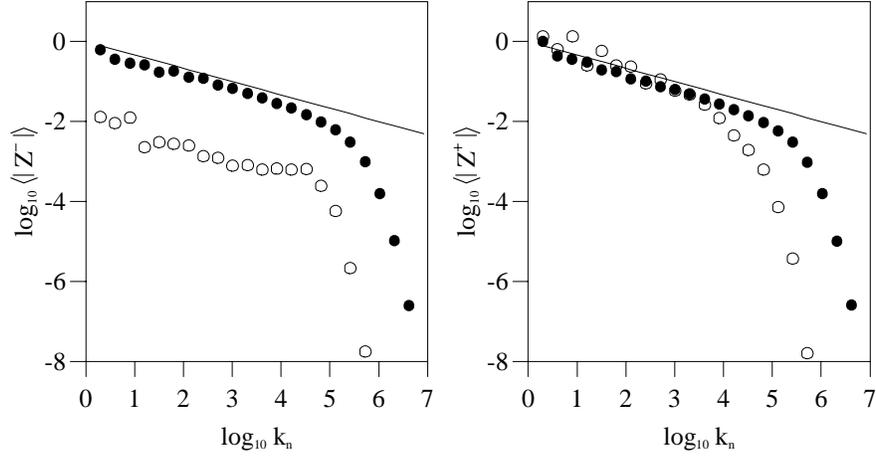}
\caption[]{{\it Left}: $\log_{10} \langle \vert Z_n^-\vert\rangle$ versus
$\log_{10} k_n$ at
different times ({\bf a}) Black circles: The average of $\vert Z_n^-\vert$
are made over the first $30$ large scale turnover times ({\bf b}) White circles:
The average is made after $300$ large scale turnover times ({\bf c}).
The straight line has slope $-1/3$. {\it Right}: The
same for $ \vert Z_n^+\vert$ 
}
\label{Zmpeps}
\end{figure}

\begin{figure}
\centering
\includegraphics[width=.9\textwidth]{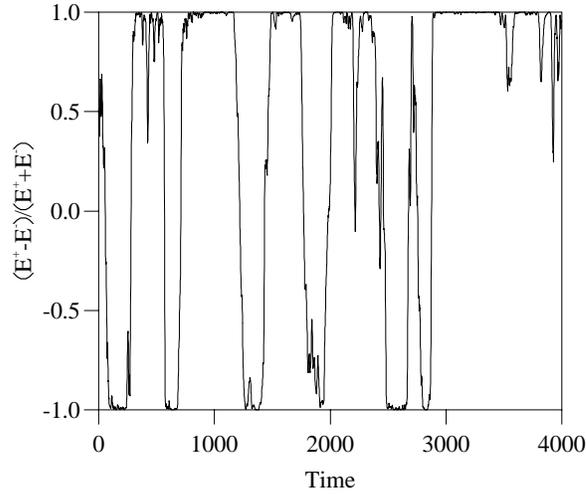}
\caption[]{Reduced cross helicity versus time in the case of an exponentially
time-correlated random gaussian forcing on the velocity variables}
\label{heleps}
\end{figure}

The main fundamental difference between hydrodynamic and MHD
shell models lies in the fact that the behaviour of the former
is not so sensitive to the type of forcing, at least
as far as the main features are concerned.
On the contrary
in the magnetic case phase space is more complex because of
the presence of invariant subspaces which can act as attractors
of the dynamics of the system, hence the type of forcing becomes
crucial in selecting the stationary state reached by the system.
The spectral properties of the 3D model have been
investigated by Frick and Sokolov in \cite{FS} under
different choices of the forcing terms. In their simulations they
observe that the spectral indexes of kinetic and magnetic
spectra depend on the level of cross helicity and magnetic helicity.
In particular spectra with spectral index $-5/3$ appear if
the cross helicity vanishes.  Even in this case results may be
deceptive. In fact, defining the reduced cross helicity $h_{R}$ as
the cross helicity divided by the total energy, long runs \cite{PeV} show
that, in case of constant forcing
on the velocity variables, even from an initial value $h_R=0$
the system evolves inevitably towards
a state in which the reduced cross helicity reaches either
the value +1 or -1,
corresponding to a complete correlation or anti-correlation
between velocity and magnetic field. In terms of attractors
the system is attracted towards one of the Alfv\`enic subspaces
where velocity and magnetic field are completely aligned or
anti-aligned. Due to the particular form of the nonlinear
interactions in MHD (\ref{nonlineare}), the nonlinear transfer
of energy towards the small scales is stopped. In this case
Kolmogorov-like spectra appear as a transient of the
global evolution. This is shown in fig.\ref{Zmpeps} where it is clearly
seen a component ($Z_n^-$) which is completely vanishing while the $Z_n^+$
spectrum becomes steeper and steeper as energy is not removed from large
scales.
If an exponentially correlated in time gaussian random forcing on the velocity
field is adopted the system
shows a very interesting behaviour. It spends long periods (several
large scale turnover times) around one of the Alfv\`enic attractors,
jumping from one to the other rather irregularly (fig. \ref{heleps}).
This behaviour
assures the existence of a flux of energy to the small scales, 
modulates the level of nonlinear interactions and the consequent
dissipation of energy at small scales, which is burstly distributed
in time. What we want to stress is
the fact that the Alfv\`enic attractors play a relevant role in the dynamics
of the system. This fact should be taken into account especially
when a stationary state is investigated in order to determine
the scaling exponents of the structure functions (see below).
Two regimes, the Kolmogorov transient and the completely aligned regime,
could be mixed during the average procedure, thus leading
to unreliable values of the scaling exponents.
\section{Fluxes, Inertial Range and Intermittency}
\begin{figure}
\centering
\includegraphics[width=.7\textwidth]{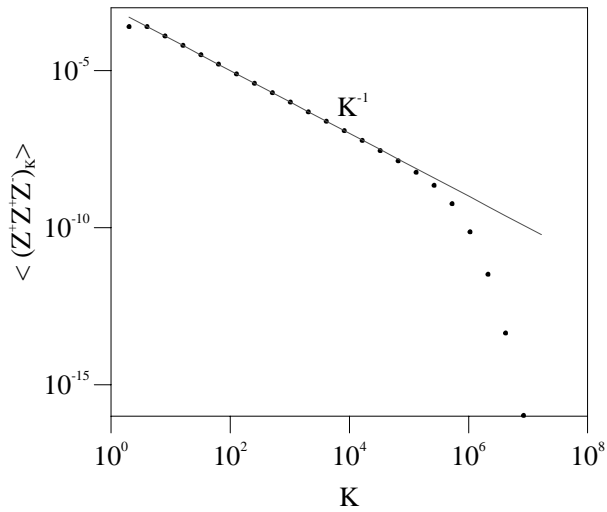}
\caption[]{Numerical check of the exact scaling relation involving
the mixed third order moment}
\label{kolmeps}
\end{figure}

The ``four-fifth" relation $\langle\,\delta v(l)^3\,\rangle=(-4/5)\epsilon l$,
where $\epsilon$ is the mean rate of energy dissipation and $l$ the separation,
derived by Kolmogorov in \cite{K41},
can be generalized to MHD flows \cite{Cha1},\cite{Cha2},\cite{PoPo}.
A corresponding relation exists in MHD shell
models, which can be derived following the considerations
in \cite{PBCFV}.
Assuming for simplicity $\nu=\eta$, the scale--by--scale energy
budget equation is:
\[
{d\over dt}\sum_{i=1,n}{|Z_{i}^{\pm}|^{2}\over 4}=
k_{n}\,(Z^{\pm}Z^{\pm}Z^{\mp})_n
-\nu\sum_{i=1,n}k_{i}^{2}{|Z_{i}^{\pm}|^{2}\over 2}+\sum_{i=1,n}{1\over 2}
\mbox{Re}\{Z_{i}^{\pm}(f_{i}^{\pm})^{\ast}\}
	\]
where the quantities $(Z^{\pm}Z^{\pm}Z^{\mp})_n$ are defined as
\begin{eqnarray}
(Z^{\pm}Z^{\pm}Z^{\mp})_n & = & {1\over 4}\mbox{Im}\{(\delta+\delta_m)\}
Z_{n}^{\pm}Z_{n+1}^{\pm}Z_{n+2}^{\mp}
+{(2-\delta-\delta_m)\over\lambda}
Z_{n-1}^{\pm}Z_{n}^{\mp}Z_{n+1}^{\pm} \nonumber \\
                          & + & 
(2-\delta-\delta_m)Z_{n}^{\pm}Z_{n+1}^{\mp}Z_{n+2}^{\pm}+
{(\delta_m-\delta)\over\lambda}Z_{n-1}^{\mp}Z_{n}^{\pm}Z_{n+1}^{\pm}\}
\end{eqnarray}
Assuming that
i) forcing terms only act at large scales;
(ii) the system tends to a statistically stationary state; (iii) in 
the infinite Reynolds numbers limit ($\nu
\rightarrow 0$) the mean energy
dissipation tends to a finite positive limit $\varepsilon^{\pm}$, we obtain 
\[
\langle\,(Z^{+}Z^{+}Z^{-})_n\,\rangle=-\varepsilon^{+}\,k_n^{-1}
\]
\[
\langle\,(Z^{-}Z^{-}Z^{+})_n\,\rangle=-\varepsilon^{-}\,k_n^{-1}
\]
These are the equations that define the inertial range of the system
and that can be easily checked and confirmed 
by numerical simulations (fig. \ref{kolmeps}). It is to be remarked that
these are the appropriate combinations that are expected to scale
exactly as $k^{-1}$.
Let us finally remind that, as far as cascade properties of shell models
are concerned, the major drawback lies in the difficulty to reproduce
cascades of quantities that are expected to flow inversely, such as
energy in 2D hydrodynamic \cite{DM} or magnetic helicity in MHD \cite{Bip}.
 
A deep understanding of intermittency in turbulence is nowdays
one of the most challenging tasks from a theoretical point of view
(see \cite{Frisch} for review).
A lot of papers have been dedicated in the last years to
investigate temporal intermittency in shell models. 
Deviations from the Kolmogorov scaling $\xi_p=p/3$ of the scaling
exponents in the structure functions,
$\langle\vert u_n \vert\rangle^p\sim k^{-\xi_p}$,
have been observed and described
in the context of a multifractal approach ~\cite{JPV}.
A precise calculation of the scaling exponents may have difficulties
related to the presence of periodic oscillations superimposed to the power
law. Another source of uncertainty is linked to the exact identification
of the inertial range where the fit should be performed.
These problems are at lenght discussed and investigated in \cite{LVov}
where a new shell model (called Sabra model) has been introduced in
the context of hydrodynamic turbulence. The Sabra model is
a slight modification of the standard GOY model and allows to
eliminate spurious oscillations in the spectra. The same problems are
in principle encountered in magnetohydrodynamic models 
thus a generalization of the Sabra model to MHD is required (\cite{Noi}).
An alternative approach to the determination of scaling exponents
for the 3D MHD GOY model
can be found in \cite{Basuetal} where concepts and techniques related to
ESS and GESS
\cite{Benzietal} are used.  

\begin{figure}
\centering
\includegraphics[width=.7\textwidth]{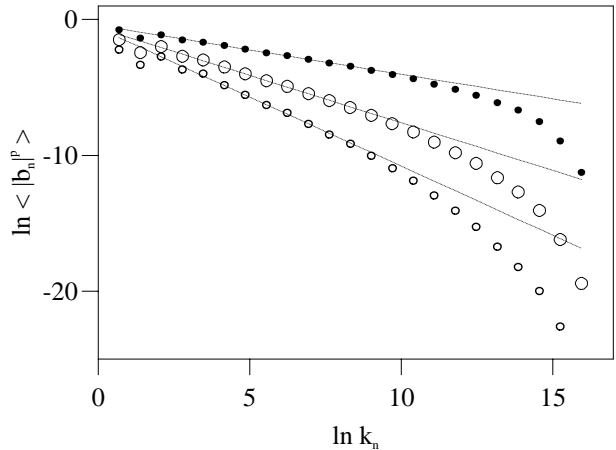}
\caption[]{Structure functions $\ln \langle\vert b_n\vert^p\rangle$
versus $\ln\,k_n$ for ({\bf a}) $p=1$ ({\it black circles})
({\bf b}) $p=2$ ({\it large white circles}) ({\bf c}) $p=3$ ({\it small white
circles}). The straight lines are the best fit in the range between $n=3$
and $n=12$
}
\label{mom123eps}
\end{figure}

We have determined the scaling exponents of the structure functions
$\langle\vert u_n \vert^p\rangle\sim k^{-\xi_p^u}$, 
$\langle\vert b_n \vert^p\rangle\sim k^{-\xi_p^u}$,
$\langle\vert Z_n^+ \vert^p\rangle\sim k^{-\xi_p^+}$,
$\langle\vert Z_n^- \vert^p\rangle\sim k^{-\xi_p^-}$
adopting a random forcing on the velocity
variables (on shell n=1 and n=2) to assure the system does not ``align".
The forcing terms were calculated solving a Langevin equation
$\dot{f_n}=-(1/\tau_0)\,f_n+\mu$, where $\tau_0$ is a correlation time
chosen equal to the large scale turnover time and $\mu$ is a gaussian
delta-correlated noise. 
The total number of shells
is $23$ and the values of viscosity and resistivity are
$\nu=0.5\cdot 10^{-9}$, $\eta=0.5\cdot10^{-9}$.
In fig. \ref{mom123eps} the first three structure functions are
plotted for the magnetic field, together with the best fit lines.
From a comparison with spectra obtained in the standard GOY model \cite{PBCFV},
it should be remarked that
the cross over region between the inertial range
and the dissipative one is not so sharp as in the hydrodynamic case.

We then decided to perform a least-square fit  
in the range, determined visually, between the shell numbers $n=3$ and $n=12$.
The values of $\xi_p^u$ and $\xi_p^b$ are reported in Table~\ref{Tab1a}
together with the
values of $\xi_p$, extracted from \cite{PBCFV}, for the
hydrodynamic GOY model. 
The values of the scaling exponents of the other structure functions
are compatible, within errors coming from the fit procedure, 
with those of the velocity variables.
It can be seen
that the values found are compatible with those obtained for the standard
hydrodynamic GOY model.
\begin{table}
\centering
\caption{Scaling exponents $\xi_p$ (GOY model), $\xi_p^u$, $\xi_p^b$}
\renewcommand{\arraystretch}{1.4}
\setlength\tabcolsep{5pt}
\begin{tabular}{llllll}
\hline\noalign{\smallskip}
$p$ & $\xi_p\;\mathrm{(GOY)}$ & $\xi_p^u$ & $\xi_p^b$
  & $$
  & $$\\
\noalign{\smallskip}
\hline
\noalign{\smallskip}
 $1$ & $    $ & $0.37 \pm 0.01$ & $0.36 \pm 0.01$  & &  \\
 $2$ & $0.72$ & $0.71 \pm 0.01$ & $0.70 \pm 0.01$   & &  \\
 $3$ & $1.04$ & $1.02 \pm 0.02$ & $1.02 \pm 0.02$   & &  \\
 $4$ & $1.34$ & $1.31 \pm 0.03$ & $1.32 \pm 0.03$   & & \\
 $5$ & $1.61$ & $1.57 \pm 0.04$ & $1.59 \pm 0.04$   & & \\
 $6$ & $1.86$ & $1.82 \pm 0.05$ & $1.84 \pm 0.05$   & & \\
 $7$ & $2.11$ & $2.05 \pm 0.06$ & $2.07 \pm 0.06$   & & \\
 $8$ & $2.32$ & $2.28 \pm 0.08$ & $2.30 \pm 0.08$   & & \\
\hline
\end{tabular}
\label{Tab1a}
\end{table}

\section{Conclusions}
In this paper we have reported about the main properties
concerning dynamical behaviour and intermittency of a shell
model for MHD turbulence.
The properties of the model reveal a complex structure of phase
space in which invariant subspaces are present. The stability properties of
the kinetic
subspace are related to a dynamo action in the system 
while Alfv\`enic subspaces act as strong attractors
which cause the system to evolve towards a state in which no energy
cascade is present. A careful choice of forcing terms seems to be
crucial in determining the stationary state reached by the system.

We want finally mention that shell models, as good candidates to
reproduce the main features of MHD turbulence, can be used to
check conjectures and ideas in astrophysical applications where
very high Reynolds numbers are often present. We briefly remind two
examples of applications. In \cite{Brand1} 
MHD shell models have been used to simulate magnetohydrodynamics
in the early universe to investigate the effects of plasma viscosity
on primordial magnetic fields. 
As second example, scaling laws found in the probability 
distribution functions of
quantities connected with solar flares (eruption events in the solar corona)
are at present matter of investigation by means of shell models.
Results on this subject can be found in \cite{BCGVV}.

I am grateful to Vincenzo Carbone, Pierluigi Veltri, Leonardo
Primavera, Guido Boffetta,
Antonio Celani and Angelo Vulpiani for useful discussions and
suggestions.

\end{document}